%% file: main.tex
\documentclass[sigconf, balance=false, prologue, table]{acmart}
\usepackage{amsmath}
\usepackage{booktabs} %
\usepackage{float}
\usepackage{layouts}
\usepackage{wrapfig}
\usepackage[ruled,vlined]{algorithm2e}
\usepackage{bbm}
\usepackage{appendix}
\input{pre}

\acmJournal{TOG}

\copyrightyear{2023} 
\acmYear{2023} 
\setcopyright{acmlicensed}\acmConference[SA Conference Papers '23]{SIGGRAPH Asia 2023 Conference Papers}{December 12--15, 2023}{Sydney, NSW, Australia}
\acmBooktitle{SIGGRAPH Asia 2023 Conference Papers (SA Conference Papers '23), December 12--15, 2023, Sydney, NSW, Australia}
\acmPrice{15.00}
\acmDOI{10.1145/3610548.3618220}
\acmISBN{979-8-4007-0315-7/23/12}

\begin{document}
\title{Subspace Mixed Finite Elements for Real-Time Heterogeneous Elastodynamics}

\author{Ty Trusty }
\email{trusty@cs.toronto.edu}
\authornote{Both authors contributed equally to this research.}
\affiliation{
\institution{University of Toronto}
\country{Canada}
}
\author{Otman Benchekroun  }
\email{otman.benchekroun@mail.utoronto.edu}
\authornotemark[1]
\affiliation{
\institution{University of Toronto}
\country{Canada}
}
\author{Eitan Grinspun}
\email{eitan@cs.toronto.edu}
\affiliation{
\institution{University of Toronto}
\country{Canada}
}
\author{Danny M. Kaufman}
\email{dannykaufman@gmail.com}
\affiliation{
\institution{University of Toronto}
\country{Canada}
}
\affiliation{
\institution{Adobe Research}
\country{U.S.A.}
}
\author{David I.W. Levin}
\email{diwlevin@cs.toronto.edu}
\affiliation{
 \institution{University of Toronto}
 \country{Canada}
}
\affiliation{
 \institution{NVIDIA}
 \country{Canada}
}




\begin{abstract}
Real-time elastodynamic solvers are well-suited for the rapid simulation of homogeneous elastic materials, with high-rates generally enabled by aggressive early termination of timestep solves. 
Unfortunately, the introduction of strong domain heterogeneities can make these solvers slow to converge.
Stopping the solve short creates visible damping artifacts and rotational errors.
To address these challenges we develop a reduced mixed finite element solver that preserves rich rotational motion, even at low-iteration regimes.
Specifically, this solver augments time-step solve optimizations with auxillary stretch degrees of freedom at mesh elements, and maintains consistency with the primary positional degrees of freedoms at mesh nodes via explicit constraints.
We make use of a Skinning Eigenmode subspace for our positional degrees of freedom. 
We accelerate integration of non-linear elastic energies with a cubature approximation, placing stretch degrees of freedom at cubature points.
Across a wide range of examples we demonstrate that this subspace is particularly well suited for heterogeneous material simulation.
Our resulting method is a subspace mixed finite element method completely decoupled from the resolution of the mesh that is well-suited for real-time simulation of heterogeneous domains.
\end{abstract}

\begin{teaserfigure}
   \centering%
  \includegraphics[width=\textwidth]{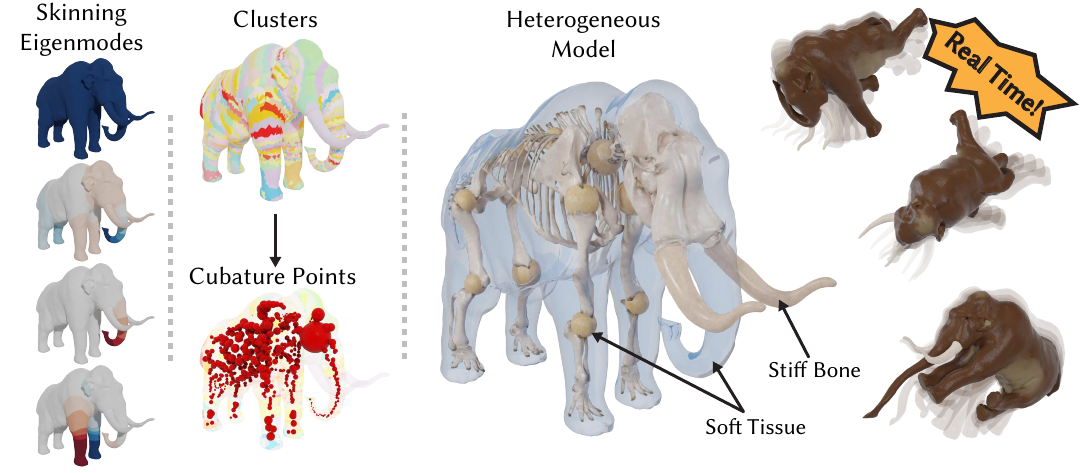} \timestamp{\tsMammothTeaser}
  \caption{%
  We propose a reduced space mixed finite element method (MFEM) built on a Skinning Eigenmode subspace and material-aware cubature scheme.  Our solver is well-suited for simulating scenes with large material and geometric heterogeneities in real-time. This mammoth geometry is composed of 98,175 vertices and 531,565 tetrahedral elements and with a heterogenous composition of widely varying materials of muscles ($E = 5 \times 10^5$ Pa), joints ($ E = 1 \times 10^5$ Pa), and bone ($E = 1 \times 10^{10}$ Pa). The resulting simulation runs at 120 frames per second (FPS).
  \label{fig:teaser-figure}
  }
\end{teaserfigure}
\maketitle

\input{intro}

\input{related}

\input{method.tex}

\input{implementation.tex}

\input{results}

\input{conclusion}

\input{acknowledgements}
\bibliographystyle{ACM-Reference-Format}

\bibliography{sample}

\input{extra_figures}

\end{document}

%% file: pre.tex
\definecolor{forestgreen}{rgb}{0.13, 0.55, 0.13}
\definecolor{lava}{rgb}{0.81, 0.06, 0.13}
\definecolor{magenta}{rgb}{0.7, 0.0, 1.0}

\newcommand{\Edit}[1]{ #1}

\definecolor{staticColor}{HTML}{005AB5}
\newcommand{\staticp}[1]{\textcolor{staticColor}{#1}}
\def\staticColorName{blue}
\definecolor{dynamicColor}{HTML}{DC3220}
\newcommand{\dynamicp}[1]{\textcolor{dynamicColor}{#1}}
\def\dynamicColorName{red}

\newcommand{\reffig}[1]{Fig.~\ref{fig:#1}}
\newcommand{\refeq}[1]{Eq.~(\ref{eq:#1})}

\newcommand*{\img}[1]{%
    \raisebox{-0\baselineskip}{%
        \includegraphics[
        keepaspectratio,
        ]{#1}%
    }%
}

\newcommand*{\timestamp}[2][-0.5cm]{%
   \sffamily \small%
  \vspace{#1}%
  \begin{flushright}%
  \protect\img{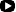}#2%
  \end{flushright}%
}
\def\tsMammothTeaser{0m06s}
\def\tsCrab{3m58s}
\def\tsBBHvssEigenmodes{2m55s}
\def\tsLSDvsEigenmodes{2m36s}

\def\tsGatormanSlingshot{4m23s}
\def\tsHeteroPendulum{4m48s}

\usepackage{amsmath}

\DeclareMathOperator*{\argmin}{argmin}

\usepackage{mfirstuc}
\MFUnocap{et}
\MFUnocap{al}

\newcommand{\W}{\boldsymbol{W}}
\newcommand{\B}{\boldsymbol{B}}
\newcommand{\0}{\boldsymbol{0}}

\newcommand{\Hw}{\boldsymbol{H}_w}
\newcommand{\Mw}{\boldsymbol{M}_w}

\newcommand{\Hx}{\boldsymbol{H}_x}

\newcommand{\fx}{\boldsymbol{f}_x}

\newcommand{\D}{\boldsymbol{D}}

\providecommand{\l}{}
\renewcommand{\l}{\boldsymbol{\lambda}}

\newcommand{\lxi}{\boldsymbol{\mu}}

\newcommand{\Gu}{\boldsymbol{G}_u}
\newcommand{\Gz}{\boldsymbol{G}_z}
\newcommand{\Hu}{\boldsymbol{H}_u}
\newcommand{\Hz}{\boldsymbol{H}_z}
\newcommand{\fu}{\boldsymbol{f}_u}

\newcommand{\s}{\boldsymbol{s}}

\newcommand{\fz}{\boldsymbol{f}_z}
\newcommand{\fxi}{\boldsymbol{f}_{\mu}}
\newcommand{\dz}{\boldsymbol{dz}}
\newcommand{\du}{\boldsymbol{du}}

\newcommand{\Rn}[1]{\mathbb{R}^{#1}}

\newcommand{\pderiv}[2]{\frac{\partial #1}{\partial #2}}
\newcommand{\pdderiv}[2]{ \frac{\partial^2 #1} {\partial^2 #2 }}

\providecommand{\s}{}
\renewcommand{\s}{\boldsymbol{s}}
\newcommand{\z}{\boldsymbol{z}}

\newcommand{\K}{\boldsymbol{K}}
\newcommand{\x}{\boldsymbol{x}}
\newcommand{\X}{\boldsymbol{X}}

\newcommand{\con}{\boldsymbol{c}}
\newcommand{\conr}{\boldsymbol{g}}

\providecommand{\u}{}
\renewcommand{\u}{\boldsymbol{u}}

\providecommand{\I}{}
\renewcommand{\I}{\boldsymbol{I}}

\newcommand{\numtets}{{|\mathcal{T}|}}
\newcommand{\numverts}{{|\mathcal{V}|}}
\newcommand{\numcubature}{{|\mathcal{C}|}}
\newcommand{\cubature}{{\mathcal{C}}}

%% file: intro.tex
\section{Introduction}
All elastic objects in the real world are heterogeneous.
Yet many of our elastodynamic simulations, especially in the \emph{real-time} regime, are evaluated on homogeneous materials. 
Applying them to
\emph{heterogeneous} materials makes these solvers slow to converge, leading to visual artifacts such as artifical damping.
These convergence artifacts are exacerbated by a strict compute-time budget; a slowly converging solve will have to be cut short as new simulation frames are demanded.
The mixed finite element method (MFEM) introduced by \citet{trusty2022mixed} shows success in preserving energetic motion for full space heterogeneous simulations.
Unfortunately, their method scales in complexity with the full mesh resolution; larger meshes quickly become unavailable for real-time simulations.
For example, the mammoth example shown in Fig. \ref{fig:teaser-figure} runs at 263 seconds per iteration (maximum 0.003 FPS), far from the common real-time target of 60 FPS.

On the other hand, subspace methods have been popular in graphics for accelerating optimization problems since \citet{pentlandwilliams1989goodvibrations}.
However, subspace methods have very well known weaknesses in representing extreme rotational motion \cite{choi2005modalwarping}, which is made worse by material or geometric heterogeneities.
Recently \citet{benchekroun2023fast} introduce Skinning Eigenmodes, a linear subspace that preserves rotation invariance during subspace simulation, and can represent rotational motion.

With the goal of simulating heterogeneous elastodynamic materials in real-time, we propose a subspace MFEM solver that makes use of a Skinning Eigenmode subspace and an accompanying heterogeneity-aware cubature approximation scheme. 
This solver inherits both the material-robust convergence benefits of its full space predecessor as well as the speed and reduced dimensionality provided by the subspace. 
The result is a convergent simulation for heterogeneous domains whose complexity is entirely decoupled from the resolution of the underlying mesh.

\begin{figure}
\includegraphics[width=\linewidth,keepaspectratio]{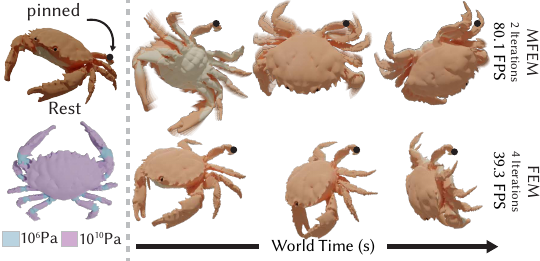}
\timestamp{\tsCrab}
\caption{A crab with a hard shell (E=1e10 Pa) and soft joints (E=1e6 Pa) is simulated with our subspace MFEM and skinning subspace FEM. With only 2 solver iterations MFEM exhibits correct rotational and elastic behavior, whereas subspace FEM with 4 iterations -- and consequently half the frame rate -- exhibits noticeable damping.
 \label{fig:crab_low_iter}}
\end{figure}

%% file: related.tex
\section{Related Work}

\subsection{Subspaces for Heterogeneous Materials}
Subspace simulation has been of interest in graphics since \citet{pentlandwilliams1989goodvibrations}, where a subspace is commonly formed via Linear Modal Analysis (LMA) of the elastic energy Hessian.  These types of modes have well known drawbacks when used for accelerating elastodynamic simulation. Specifically, they struggle representing large non-linear deformations, such as rotations, which are a salient feature of heterogeneous material simulation. 

\citet{barbicjames2005realtimestvk} use modal derivatives, whose aim is to supplement primary LMA modes with higher quality derivative modes to help the subspace stay up to date with the current deformation. Unfortunately, modal derivatives  do not perfectly span rotations (See \reffig{skinning_eigenmodes_vs_modal_derivs}). Modal warping \cite{choi2005modalwarping}, Rotation Strain coordinates \cite{huang2011rotationstrain},  sub-structuring \cite{barbic2011substructuring, kimjames2011substructuring} and rigid-frame embedding \cite{james2002dyrt, terzopoulous1988rigidanddeformablecomponents}  all mitigate this issue by factoring out rotational motion and keeping track of it separately. These methods unfortunately scale in complexity with the number of rotations to be tracked, of which there may be many in a large-scale heterogeneous material. \Edit{While non-linear subspaces via Deep Neural Networks have also been proposed \cite{shen2021auto}, the resulting complexity of the subspace requires many optimization steps in order to reach a solution \cite{sharp2023data}.}

\begin{figure}
\includegraphics[width=\linewidth,keepaspectratio]{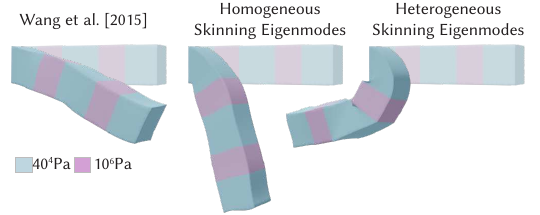}
\timestamp{\tsLSDvsEigenmodes}
\caption{Material sensitive skinning modes directly lead to richer motion for heterogeneous materials.}
\label{fig:skinning_eigenmodes_vs_bbw}
\end{figure}

Another option is to use linear skinning subspaces, which can represent rotations implicitly upon their construction. Many skinning subspaces are chosen to represent smooth, local deformations
\cite{lei2020medialelastics, jacobson2012fast, wang2015linearsubspacedesign, brandt2018hyperreduced}. Smoothness, however, is not an optimal prior when a material has sharp transitions in material properties (See \reffig{beam_twist} and \reffig{skinning_eigenmodes_vs_bbw}).
\citet{faure2011sparsemeshless} describe a local, material-sensitive set of skinning weights.  Their construction rely on additional user parameters to control the smoothness of the subspace. 
\citet{benchekroun2023fast} propose Skinning Eigenmodes, a method of constructing globally supported skinning weights that reflect material properties from an eigendecomposition of the elastic energy Laplacian.
The globality of this subspace allows for a compact representation of fine scale motion. 
While material aware, Skinning Eigenmodes in standard finite element solvers still suffer degraded convergence with large heterogeneities. We show that the combination of this subspace with a mixed finite element method is the key to robust real-time heterogenous simulation.

\subsection{Fast Elastic Solvers for Heterogeneous Materials}
Standard discretizations struggle when applied to heterogeneous elastodynamics problems which motivates our use of a mixed discretization ~\citep{trusty2022mixed}. Typically, work on efficient simulation of heterogenous materials is centered around homogenization or numerical coarsening~\citep{kharevychCoarsening2009,chenCoarsening2015,chenCoarsening2017} which uses a coarse (lower than material assignment resolution) mesh as a reduced space and homogenizes material properties within each coarse element. However these methods only simulate aggregate material behavior -- by construction they cannot accurately represent the heterogeneous strains induced by material or geometric heterogeneity as the shape functions themselves are typically polynomial within each element. \Edit{While \citet{chenCoarsening2018} derive material-adaptive multi-resolution basis functions, their approach depends on a non-physical Rotation-Strain post-warping effect using Rotation Strain coordinates, which they show produces artifacts.}  Rather, our material-aware skinning weights allow more visually exacting reconstruction of animated motion using fewer degrees-of-freedom.

Another common method for accelerating non-linear elastic PDEs discretized via Finite Elements is by a cubature approximation \cite{an2008optimizingcubature, vonTycowicz2013efficientrealtime} of the elastic energy. 
This approximates the total elastic energy with reweighed contributions from a set of sparsely sampled representative tetrahedra.
The computation of these cubature points and weights is done in an expensive offline training phase, requiring the user to provide data with prior knowledge of the deformations they expect to encounter at run-time.
Instead, \citet{jacobson2012fast} accelerate an elastostatic solver by allowing 
tetrahedra to share strain quantities with other tetrahedra in their cluster.
These clusters are found efficiently via a $k$-means clustering on the  skinning weights, allowing the clusters to reflect the properties of the skinning weights.
We combine both approaches: we find a strong set of cubature points as the centroid of the $k$-means clusters without requiring a training phase.

%% file: method.tex
\begin{figure*}[h]
\includegraphics[width=\textwidth,keepaspectratio]{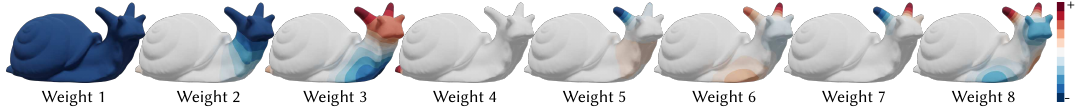}
\caption{ The skinning weights we get from Skinning Eigenmodes are naturally material aware. High frequency modes are concentrated on soft parts of the snail, which are more likely to exhibit rich deformation. In contrast, the stiff shell only has access to a constant skinning weight (shared by all parts of the snail), allowing rigid motion to be producible within our skinning subspace.
\label{fig:snail_skinning_subspace}}
\end{figure*}

\section{Full Space Mixed FEM}

Our starting point is the mixed finite element method (MFEM) of \citet{trusty2022mixed}. We discretize the domain with a tetrahedral mesh with $\numverts$ vertices and $\numtets$ elements.

We store positions as the coefficients $\x \in \Rn{3 \numverts  \times 1}$ of Lagrange finite elements.
We then introduce stretch degrees of freedom (DOFs)  $\s \in \Rn{ 6 \numtets \times 1} $, corresponding to the symmetric stretch component of the polar decomposition of the deformation gradient ($F = RS$).

We maintain consistency between our positional and stretch DOFs $\con(\x, \s) = \D (\bar{\s}(\x) - \s)$, where $\bar{\s}$ evaluates the stretch at each element as a function of $\x$. We make use of $D = \mathrm{diag}([ 1 \; 1 \; 1 \; 2 \; 2 \; 2])$ and $\D = \I_{\numtets}  \otimes D$ to account for the symmetric off-diagonal terms in $S$.

This leads to the MFEM elastodynamic optimization problem,
\begin{align}
\x^*, \s^*, \l^*  =  \argmin_{\x, \s} \max_{\l}   \Psi_x(x) + \Psi_s(s) + \l^T \con(\x,\s)  \;
\label{eq:mfem-discrete-optimization}
\end{align}
where $ \Psi_{\x}(\x)$ is the \emph{quadratic} component of the elastodynamic energy that depends only on positional DOFs, $ \Psi_{\s}(\s)$  is the elastic strain energy written in terms of the stretch DOFs and the last term enforces the consistency constraint with Lagrange multipliers $\l \in \Rn{6 \numtets }$.
The solution is characterized by the  KKT optimality conditions, which can be solved via a Newton-type method \cite{trusty2022mixed}.

\section{Subspace Mixed FEM}
We introduce a linear subspace $\boldsymbol{B} \in \Rn{3 \numverts  \times r}$ for our positional DOFs, and approximate them with $\x  \approx \B \u$, where $\u \in \Rn{r}$, $r \ll 3\numverts $, are subspace coefficients. 
With some precomputations, this subspace can be used to evaluate the quadratic $\Psi_x(\cdot)$ at run-time explicitly in terms of reduced dimensions. 
By contrast, fast evaluation of the  non-linear stretch energy, $\Psi_s(\cdot)$, and the consistency constraint term requires the use of numerical cubature \cite{an2008optimizingcubature}.
Evaluating these corresponding energy densities over a subset of all tetrahedra, $\cubature$, and reweighing their contributions according to a precomputed cubature weight yields
\begin{align}
    &\Psi_s(\s)  
     \approx \sum_c^{\numcubature} w_c \psi_z(\z_c) = \Psi_z(\z), \\  
    &\l^T  \con(\x,\s)  \approx \sum_c^{\numcubature}  w_c \lxi_c ^T D (\bar{\z}_c(\u) - \z_c) = \lxi^T \conr(\u,\z) %
\end{align}

where we have introduced $\z \in \Rn{6 \numcubature}$, the stretch DOFs at the cubature tetrahedra,
and $\lxi \in \Rn{6 \numcubature }$, the Lagrange multipliers enforcing the consistency constraint at the cubature points.

We can finally rewrite the optimization problem entirely in terms of reduced space DOFs:
\begin{align}
\u^*, \z^*, \lxi^*  =  \argmin_{\u, \z}\max_{\lxi}   \Psi_u(\u) + \Psi_z(\z) + \lxi^T \conr(\u,\z)
\label{eq:mfem-discrete-optimization-reduced}
\end{align}
We solve this optimization using Sequential Quadratic Programming (SQP), where search directions for the $(k+1)$-th iteration are found by solving the KKT system
\begin{align}
        \begin{bmatrix}
        \Hu & \0 & \Gu^T \\
        \0  & \Hz & \Gz \\
        \Gu & \Gz & \0
        \end{bmatrix}
        \begin{bmatrix}
        \du \\ \dz \\ \lxi
        \end{bmatrix}
        = -
        \begin{bmatrix}
        \fu \\ \fz \\ \fxi
        \end{bmatrix},
\end{align}
where all quantities are evaluated using DOFs from the previous iteration, $\{\u^k, \; \z^k\}$. $\Hu = \B^T\Hx \B \in \Rn{ r \times r }$ 
and $\Hz = \pdderiv{\Psi_z}{\z} \in \Rn{6 \numcubature \times 6 \numcubature}$ are the reduced Hessians \Edit{(with $\Hx$ being the full-space Hessian)};
$\fu = \B^T\fx \in \Rn{r }$, 
$\fz = \pderiv{\Psi_z}{\z} \in \Rn{6 \numcubature }$, and $\fxi = \conr(\u^k,\z^k) \in \Rn{6 \numcubature}$ are the reduced forces \Edit{(with $\fx$ being the full-space force)};
$\Gu = \pderiv{\conr}{\u} \in \Rn{6 \numcubature \times r}$ and $\Gz = \pderiv{\conr}{\z} \in \Rn{6 \numcubature \times 6 \numcubature} $ are the reduced space constraint Jacobians.
The transpose is omitted from $\Gz$ since it is a diagonal matrix of cubature weights.

We condense this system by applying a series of Schur complements so that for $\du$ we instead solve
\begin{equation}
    (\Hu + \K) \du = -\fu + \Gu^T \Gz^{-1}(\fz - \Hz \Gz^{-1} \fxi), 
\label{eq:descent-direction}
\end{equation}
where $\K = \Gu \Gz^{-1} \Hz \Gz^{-1} \Gu^T$, %
and for $\dz$ and $\lxi$ we solve
\begin{align}
\dz &= -\Gz^{-1}(\fxi + \Gu \du),
\label{eq:subspace_mfem_stretch}  \\
\lxi &= -\Gz^{-1}(\fz + \Hz \dz).
\label{eq:subspace_mfem_lagrange}
\end{align}
The updates for the next SQP iteration are then $\u^{k+1} = \u^k + \alpha \du$ and $\z^{k+1} = \z^k + \alpha \dz$, where $\alpha$ is a step size given by backtracking line search over the Lagrangian, $\mathcal{L}(\u,\z,\lxi) = \Psi_u(\u) + \Psi_z(\z) + \lxi^T \conr(\u,\z)$. In this final form, none of the terms depend on a full space quantity, so the update for $\du$ is efficiently solved with a direct dense linear solver, and the updates for $\dz$ and $\lxi$ are local and performed in parallel, making their cost negligible.

\begin{figure}
\includegraphics[width=\linewidth,keepaspectratio]{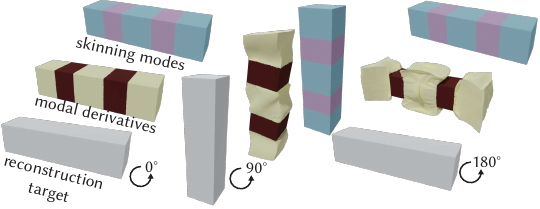}
\caption{Modal derivatives are not suited for reconstructing rotations on the input shape. Fixing these artifacts typically requires explicitly tracking a rigid frame 
\cite{terzopoulous1988rigidanddeformablecomponents}. }
\label{fig:skinning_eigenmodes_vs_modal_derivs}
\end{figure}

\section{Subspace Construction}
\subsection{Skinning Eigenmode Subspace}
There are many ways to construct our positional subspace $\B$. We opt for using a skinning subspace \cite{hahn2012rigspacephysics, brandt2018hyperreduced}. As discussed by
\citet{benchekroun2023fast}, these subspaces span rotations (as shown in \reffig{skinning_eigenmodes_vs_modal_derivs}), a particularly salient feature for heterogeneous stiff materials.

Other linear subspaces, such as modal derivatives \cite{barbicjames2005realtimestvk}, do not generally span rotations~\cite{benchekroun2023fast}. For free flying motion, this limitation
may be addressed by embedding a rigid frame that is tracked explicitly during the simulation \cite{terzopoulous1988rigidanddeformablecomponents}. For heterogeneous materials with multiple independent stiff components that do not necessarily rotate in unison (such as the bar in \reffig{skinning_eigenmodes_vs_modal_derivs}), keeping track of potentially many rigid frames becomes increasingly inconvenient, and scales in complexity with the heterogeneity of the material.

We opt to build our skinning subspace using Skinning Eigenmodes  \cite{benchekroun2023fast}, which span rotations and furthermore provide a straightforward automatic method to generating material-aware skinning weights (see \reffig{snail_skinning_subspace}).  
Specifically, we obtain a set of skinning weights $\W \in \Rn{\numverts \times m} $ by solving the weight space generalized eigenvalue problem,
\begin{align}
    \Hw \W = \Mw \W \boldsymbol{\Gamma}.
    \label{eq:gevp-skinning-eigenmodes}
\end{align}
Above, \Edit{$ \Hw = \pdderiv{\Psi}{\x_1} +\pdderiv{\Psi}{\x_2} + \pdderiv{\Psi}{\x_3}  \in \Rn{\numverts \times \numverts } $} is the elastic energy Laplacian (subscripts $\{1, 2, 3\}$ denote each of the 3 dimensions) and $\Mw \in \Rn{\numverts \times \numverts }$ is the scalar mass matrix.
The use of the elastic energy Laplacian is what provides this subspace with its material-aware properties.

\Edit{The decomposition provides us with eigenvalues $\boldsymbol{\Gamma}$ and eigenvectors $\W$, the latter of which correspond to linear blend skinning weights.}
The use of the elastic energy Laplacian is what provides this subspace with its material-aware properties.
These skinning weights $\W$ can then be used to construct our subspace basis $\B$ using the standard linear blend skinning Jacobian formula, 
\begin{align}
    \B = \I_{3}  \otimes (  ( \boldsymbol{1}_m^T \otimes \bar{\X} ) \odot  ( \W \otimes  \boldsymbol{1}_{4 }^T )),
\end{align}
where $\bar{\X} \in \Rn{ \numverts \times 4}$ are the rest positions in homogeneous coordinates.
We can relate $m$, the number of skinning weights to our subspace DOFs via $r = 12 m$.

\subsection{Cubature Construction}

\begin{figure}
\includegraphics[width=\linewidth,keepaspectratio]{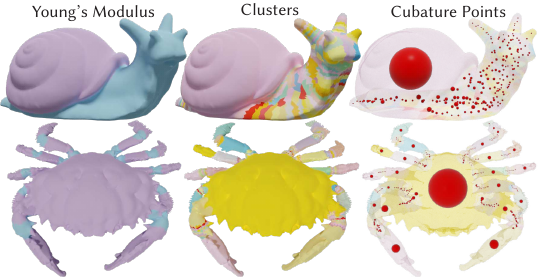}
\caption{ Our cubature points are found as the centroids of each k-means cluster. Note that our centroids are sensitive to the heterogeneity of the Young's modulus. Stiffer regions can have their strain be approximated with fewer cubature points.}
 \label{fig:cubature_vis}
\end{figure}

A cubature scheme is characterized by a set of cubature points, and their corresponding cubature weights.
A good cubature scheme is crucial for the quality of our subspace approximation.
Undersampling leads to spurious deformations \cite{mcadams2011efficient},
whereas excessive sampling introduces unnecessary cost.

A cubature scheme can be optimized to fit a training dataset ~\cite{an2008optimizingcubature}.
Cubature weights are computed via a Non-Negative Least Squares (NNLS) fitting of the forces observed in the training set.
Cubature points are then greedily added at elements where the current cubature fitting most poorly reconstructs the training forces.
This approach is well suited for scenarios where the user knows a priori the types of deformation they want to approximate.
However, for scenarios where a user is exploring deformations for potentially many meshes at a time, the requirement of building a good cubature training set as well as the time it takes to iteratively solve a large NNLS problem for the cubature weights can overly constrain the creative process. 
\Edit{\citet{yang2015Expediting} propose an avenue for acceleration based on a Preconditioned Conjugate Gradient method.}

We propose an alternative, fast and simpler method for constructing our cubature approximation that is well suited for heterogeneous materials, inspired by the clustering scheme of \citet{jacobson2012fast}.
To sample cubature tetrahedra, we cluster our domain and choose the tetrahedra closest to the centroid of each cluster. 
We construct these clusters from a $k$-means clustering on our skinning weights.
\begin{align}
    l = \textrm{kmeans}(\W_{\mathcal{T}}{\boldsymbol{\Gamma}}^{-2}, \numcubature  ),
\end{align}
where $\W_{\mathcal{T}}$ are our skinning weights averaged from the vertices to the elements. 
\Edit{We weigh each skinning weight by its inverse squared eigenvalue $\boldsymbol{\Gamma}^{-2}$ in order to favor weights that correspond to low energy deformations, which are more likely to occur at run-time.} We then compute cluster centroids and choose our cubature points as the tets closest to each centroid.
The cubature weights are then trivially computed as the mass of each cluster. 
Using the skinning weights as our clustering features allows the cubature scheme to reflect the properties of our skinning weights, such as material and geometric heterogeneity, or any pinning constraints that may have been imposed on our skinning weights. 
In particular, note from \Edit{\reffig{cubature_vis} and \reffig{hetero_geometry_cubature_points}}  that our cubature sampling parallels the anticipated strain
heterogeneity of the domain: regions more likely to deform, such as soft regions or thin regions, will be sampled relatively densely; regions less likely to deform, such as stiff or thick regions, will be sampled relatively sparsely, as shown in \reffig{hetero_geometry_cubature_points}.

%% file: implementation.tex
\begin{figure}
\includegraphics[width=\linewidth,keepaspectratio]{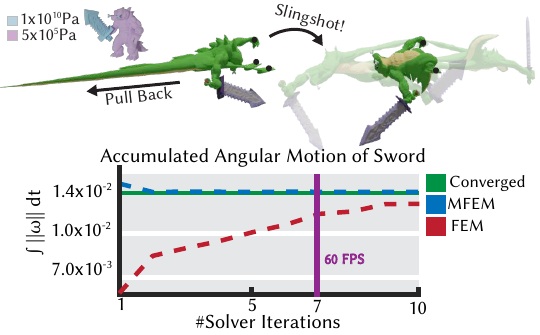}
\timestamp{\tsGatormanSlingshot}
\caption{ Our subspace MFEM solver almost perfectly reproduces the angular motion of the sword over the first 25 simulation timesteps, whereas FEM consistently underestimates it. 
 \label{fig:gatorman_slingshot}}
\end{figure}

\begin{figure}
\includegraphics[width=\linewidth,keepaspectratio]{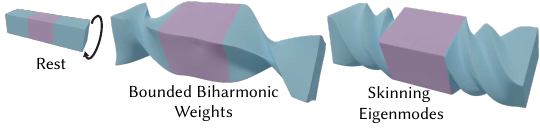}
\timestamp{\tsBBHvssEigenmodes}
\caption{
Smooth local skinning weights (left), such as Bounded Biharmonic Weights \cite{jacobson2011boundedbiharmonicweights} are not optimal for modeling materials with sharp heterogeneities. Skinning Eigenmodes (right) are material sensitive and lead to a sharp resolution of extreme twisting motions.  \Edit{Both simulations use MFEM}.
 \label{fig:beam_twist}}
\end{figure}

\section{Implementation}
\label{sec:implementation}
We implement our method in both Matlab and C++, with geometry processing utilities provided by libigl \cite{libigl} and gptoolbox \cite{gptoolbox} and physics utilities provided by Bartels \cite{bartels}. Our C++ code is parallelized with OpenMP \cite{chandra2001parallel}. For modelling and rendering we use Blender \cite{blender}. To solve the Generalized Eigenvalue Problem in \refeq{gevp-skinning-eigenmodes}, as well as the $k$-means clustering, we use Matlab's $\mathrm{eigs}()$ and $\mathrm{kmeans}()$ functions. \Edit{To solve for the search direction (\refeq{descent-direction}) in each Newton iteration we use Eigen's \cite{eigenweb} SimplicialLLT} direct solver.

Algorithm \ref{alg:simulationStep} provides pseudocode for a single simulation step of our subspace Mixed-FEM solver. 

Matrices coloured in \staticp{\staticColorName} remain constant throughout the simulation, whereas matrices in \dynamicp{\dynamicColorName} change every timestep, but remain fixed throughout Newton iterations. 

At the end of each simulation step, mesh geometry is usually queried for visualization purposes. 
Standard subspaces require full space projection, $\x = \B \u$, to in order to capture the deformed mesh geometry. 
This is a full space operation that can easily become the bottleneck for any subspace simulation application.
Instead, we perform this step entirely on the GPU \cite{barbicjames2005realtimestvk}. 
As our subspace is a skinning subspace, it's especially convenient to perform this step in the vertex shader \cite{benchekroun2023fast}:
we pass the skinning weights forming our subspace $\W$ to our vertex shader as vertex attributes in a preprocessing step, and send our reduced space coordinates $\u$ as uniforms each draw call.
As Table \ref{table:timings_mfem} shows, this effectively makes the computation time for this step negligible compared to the other stages of our pipeline.

\begin{algorithm}[h]
   \caption{Performs one simulation step of our subspace Mixed-FEM solver \label{alg:simulationStep}}
 \SetKwFunction{FsimulationStep}{simulationStep}
  \SetKwProg{Fn}{Function}{:}{}
\Fn{\FsimulationStep{$\u, \z$}}{
\While{$\mathrm{not\, converged}$}
{ 
$\Hz, \staticp{\Hu} \gets \mathrm{hessians}(\u, \z)$ \\
$\fz, \dynamicp{\fu}, \fxi \gets \mathrm{gradients}(\u, \z)$ \\
$\staticp{\Gz}, \Gu \gets  \mathrm{constraintGradients}(\u, \z)$ \\ 
$\K \gets \Gu \staticp{\Gz^{-1}} \Hz \staticp{\Gz^{-1}} \Gu^T$   // assemble stiffness matrix \linebreak \\

// Global linear solve \\
$\du \gets (\staticp{\Hu} + \K)^{-1} (\Gu^T \staticp{\Gz^{-1}}(\fz - \Hz  \staticp{\Gz^{-1}} \fxi) - \dynamicp{\fu})$ \\

// Local solves \\
$\dz = -\staticp{\Gz^{-1}}(\fxi + \Gu \du)$   \\
$\lxi = -\staticp{\Gz^{-1}}(\fz + \Hz \dz) $   \linebreak \\

$\alpha \gets \mathrm{lineSearch}(d\u, d\z, \lxi)$ \\
$ \u  \gets \u + \alpha  d\u $ \\
$ \z  \gets \z + \alpha  d\z $ \\
}
\KwRet $\u , \z$
}
\end{algorithm}

%% file: results.tex
\section{Results \& Discussion}

In the following examples, without loss of generality, we apply implicit Euler time stepping and use the fixed corotational (FCR) elasticity model \citep{fcr} \Edit{(any hyperelastic model is applicable \reffig{material_models})}. MFEM denotes our subspace MFEM solver and FEM denotes a solver which uses the same skinning subspace, but applied in standard FEM. 

\subsection{Iteration Ablation}
The advantages of our subspace MFEM solver become especially apparent for truncated real-time simulations with large heterogeneities. \reffig{crab_low_iter} shows a crab model with a stiff shell and soft joints pinned at one of its hind legs and falling under gravity.
The subspace simulation is carried out with 16 skinning modes and 342 cubature points. 
We allow only two solver iterations every timestep and compare results between our subspace MFEM solver, and a traditional subspace FEM solver.
The FEM example manifests a very common solver truncation artifact which \emph{heavily} damps motion.
By contrast, our MFEM solver easily allows the crab to exhibit rich rigid motion.

\begin{table*}[hbt]
\rowcolors{2}{white}{cyan!25}
\caption{We report average times (in milliseconds) for one iteration of subspace MFEM/FEM and full-space simulations for meshes of various complexity. \textbf{MFEM} corresponds to a simulation step time for our subspace mixed FEM solver, \textbf{FEM} is the time for a subspace FEM solve step, and $m$ and $\numcubature$ are, respectively, the number of skinning modes and cubature points used in both subspace solvers. \textbf{Proj} is the time for the full-space projection used in the subspace solvers. Lastly, \textbf{Full MFEM} is the time for a full-space MFEM iteration (\citet{trusty2022mixed}).
\label{table:timings_mfem}}
\begin{tabular}{c|c|c|c|c|c|c|c|c}
      \textbf{Mesh} & $\numverts$ & $\numtets$ & m &  $\numcubature$  & \textbf{MFEM} (ms) & \textbf{FEM} (ms) & \textbf{Proj} (ms)  & \textbf{Full MFEM} (ms) \newline \newline \\
      Octobot (\reffig{grad_convergence}) &  32 591  & 132 124   & 5 & 227  & 1.19 & 1.10 & 0.42   & 3,099.1 \\
      Gatorman (\reffig{gatorman_slingshot})& 54, 235   &   227, 035 & 10 & 192 & 2.01 & 2.04 & 0.41 & 11,442.7 \\
      Mammoth (\reffig{teaser-figure})& 98, 175 &  531, 565 & 16 &  581 & 7.37 & 7.56 &  0.54 & 263,545 \\
     Crab (\reffig{crab_low_iter})&  57,529 & 223, 565 & 16 & 342 &    5.87 & 5.51  & 0.49  & 7,483.75\\
\end{tabular}
\end{table*}

\begin{figure}
\includegraphics[width=\linewidth,keepaspectratio]{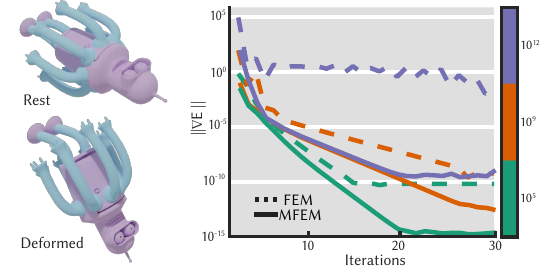}
\caption{ We start the octobot mesh in the deformed state (bottom left). We then run 30 iterations of FEM and MFEM iterations as we vary the Young's modulus of the stiff region (shown in purple). The soft region (shown in blue) remains fixed at $1\times 10^5 Pa$. For large material heterogeneities, FEM takes much longer to converge than our MFEM.
 \label{fig:grad_convergence}}
\end{figure}

\subsection{Complex Deformation}
As shown in \reffig{beam_twist}, our subspace solver can reproduce extreme twisting motions for a heterogeneous candy with a hard middle ($1\times10^{10}$ Pa) and soft extremities ($1\times10^6$ Pa). 
The twist is enforced via a spring force, and the whole simulation is carried out in a subspace of 16 skinning weights and 192 cubature points.
We compare our subspace's result to one created with bounded biharmonic weights \cite{jacobson2011boundedbiharmonicweights, lei2020medialelastics} with weight handles located about samples found via farthest point sampling. 
While bounded biharmonic weights provide a smooth basis for simulation, this subspace is not aware of the heterogeneity present within the candy's domain, resulting in most of the modes locking their motions to ensure the stiff regions remain undeformed, leaving little degrees of freedom available to accomodate the twist. 
By contrast, our skinning eigenmode subspace is sensitive to the material properties of the candy, which allows our simulation to better capture this sharp transition in the material properties of the domain.

\subsection{Material Heterogeneity}
\reffig{grad_convergence} investigates how the extent of the heterogeneity affects the convergence of our subspace Mixed-FEM solve and specifically compares against the convergence of a traditional FEM simulation.

We start with the Octobot mesh in a deformed state and run both subspace FEM and MFEM solvers for a single timestep. 
The subspace used is composed of 16 skinning modes and 800 cubature points.  
We perform this experiment for 3 different Young's moduli and plot the iteration progression of the Newton decrement for each solve.

\reffig{teaser-figure} stress tests our solver's ability to simulate large-scale models with a high number of material discontinuities.
Here, a Mammoth with stiff skeleton bones ($1 \times 10^{10}$ Pa), soft joints ($5 \times 10^5$ Pa), and softer muscle ($1 \times 10^5$ Pa) is excited by an external periodic force applied on its back bone, moving it up and down and thrashing it around the scene.
We observe energetic rag-doll rotational motion of the limbs and body, a detail noticeably absent from the unconverged subspace FEM simulation provided in the supplemental video.

\subsection{Geometric Heterogeneity}
Heterogenity of elastic moduli is just one possible source of 
large variations in elemental strains. Another possible source is
the \emph{geometry} of the domain; heterogeneous
thickness, for instance, can lead to comparatively small and large strains
in slender and thick regions, respectively (see \reffig{hetero_geometry}). 

In this example, we wind up a pendulum, twist it back a few times, and release it, allowing it to unwind and come to rest.
We carry out the simulation in a subspace composed of 16 Skinning Modes and 400 cubature points (\reffig{hetero_geometry_cubature_points}).
Starting both methods at the twisted state, we simulate the unwinding with MFEM and FEM with one iteration per timestep. We observe that MFEM maintains the same energy preserving benefits, while FEM again exhibits rotation damping artifacts. 
This example uses homogeneous material properties, emphasizing that our method
offers an advantage when the \emph{strain} is heterogeneous, whether induced 
by constitutive or geometric properties.

\begin{figure}
\includegraphics[width=\linewidth,keepaspectratio]{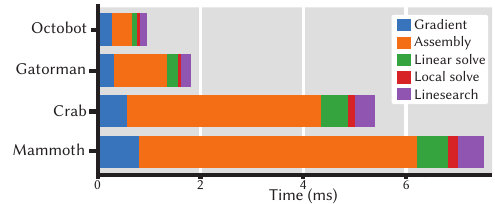}
\caption{A timing breakdown of the core components for a single subspace MFEM simulation step of the Octobot (\reffig{grad_convergence}), Gatorman (\reffig{gatorman_slingshot}), Crab (\reffig{crab_low_iter}), and Mammoth (\reffig{teaser-figure}) simulations.
\label{fig:timing_breakdown}} 
\end{figure}

\subsection{Mode-Cubature Pareto Fronts}
In \reffig{pareto_search}, we investigate how both our subspace parameters, the number of skinning modes and the number of cubature points, affect both the accuracy of our subspace approximation as well as the time it takes to run.
With the octobot starting in the bent position shown in \Edit{\reffig{grad_convergence}} and allowing one simulation timestep to occur.
We carry out this experiment assuming a homogeneous material with a Young's Modulus of $10^5$ Pa.

To measure the accuracy of the converged solution (\reffig{pareto_search}, left), we project our subspace solution back to the full space and evaluate the gradient of the full space elastodynamic optimization problem, which should be $0$ for an accurate, converged result. 
In particular, note that the top left part of the grid-search makes use of many skinning modes, but still incurs a lot of error. 
This may seem unintuitive, but in fact stems from the introduction of a null space in our cubature approximation. 
Because we have so few cubature points in this regime, but many degrees of freedom for motion, spurious 0-energy oscillations manifest, a known cubature pitfall \cite{mcadams2011efficient}.
In practice, we've found that setting the number of cubature points to $20 \times$ the number of skinning modes allows us to safely steer clear of this regime \Edit{(\reffig{cubature_limitations})}, and all the examples reported in the rest of this paper do not exhibit these spurious deformations.

\subsection{Timing Comparison and Discussion}
\begin{figure}[h]
\includegraphics[width=\linewidth,keepaspectratio]{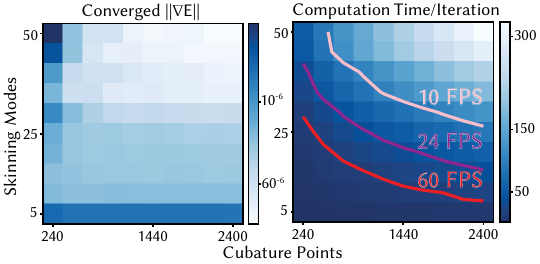}
\caption{Pareto-search exploring the cost/benefit tradeoff of varying our two subspace parameters, the number of skinning modes $m$ and the number of cubature points $\numcubature$ . We visualize the resulting full space energy gradient after convergence (left) as well as the computation time per newton iteration (right).
 \label{fig:pareto_search}}
\end{figure}

Table \ref{table:timings_mfem}  provides timings per MFEM and FEM iteration respectively.
We compare timings for the Octobot (\reffig{grad_convergence} and \reffig{pareto_search}), the Gatorman (\reffig{gatorman_slingshot}), the Mammoth (\reffig{teaser-figure}) and the Crab (\reffig{crab_low_iter}) with varying subspace sizes. We also compare our solver's performance to the full-space MFEM solver of \citet{trusty2022mixed} and attain an average speed up of over 3 orders of magnitude. 
The additional computation required of our MFEM solver, when compared to FEM, is only the local stretch and Lagrange multiplier solves \refeq{subspace_mfem_stretch}, which incurs an added $O(k)$ operations. 
This step only incurs a marginal difference as shown clearly in the timing breakdown of \reffig{timing_breakdown}, which shows the MFEM simulation time is dominated by the $O(m^2k)$ dense $\K$ matrix assembly.

With an asymptotically equivalent runtime as FEM, as well as more favorable energetic behavior at low-iterations, our solver enables real-time heterogenous domain simulation. In contrast, an equivalent subspace size requires FEM to perform  more iterations (\reffig{crab_low_iter} and \reffig{gatorman_slingshot}), making real-time simulation unattainable in many cases.%

\subsection{Artifact Tradeoffs between MFEM and FEM}
While FEM exhibits extreme damping artifacts at low iteration counts, our solver can exhibit overly-energetic motion at low-iteration counts.
\reffig{gatorman_slingshot} shows a subspace simulation on a case with extreme deformation and a localized external force.

Here, a soft gatorman ($5\times 10^{5}$ Pa) wielding a stiff ($1\times10^{12}$ Pa) sword is pulled back from its tail (using a soft penalty constraint) and slingshotted towards its enemies. 
The subspace for the simulation is composed of 10 skinning modes and 192 cubature points.
We compare our subspace simulation results with those of a traditional FEM solver as we increase the number of solver iterations. 
Note that our subspace allows us to capture the localized rotational motion of the sword which is absent from the FEM solution.
We find that MFEM is overly energetic at low iterations, causing an initial overestimation of angular motion. This results in jittering artifacts, which quickly disappear when taking more than one solver iteration.
In contrast, FEM requires many more iterations to recover the correct rotational behavior (see the supplemental video for a demonstration).

\subsection{Limitations of Global Subspaces}
\begin{figure}
    \centering  \includegraphics{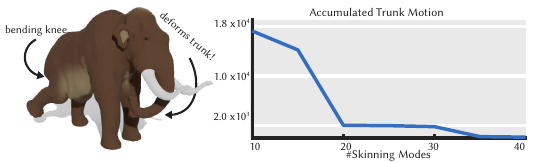}
    \caption{Because of our globally supported subspace, bending the knee of the mammoth causes the trunk to deform. This artifact goes away as we increase the size of the subspace.}
    \label{fig:global_limitations}
\end{figure}

The global support of the skinning eigenmode subspace allows our simulation to efficiently capture complex ranges of motion. 
This can lead to noticeable global artifacts when a user is exciting a local region of the mesh. 
For example \reffig{global_limitations} shows a user bending the mammoth's hind leg, causing a jerk motion in the mammoth's trunk.
We measure the amount of deformation induced by summing accumulated vertex displacements throughout the simulation. 
We increase the size of the subspace and observe that this artifact goes away as the number of skinning modes increases.

%% file: conclusion.tex
\section{Conclusion and Future Work}
We have presented a new subspace mixed finite element method that offers real-time elastodynamic simulation for large-scale heterogeneous domains.
Typical subspace methods experience degraded performance and jarring artifacts in these settings. 
We show that coupling a skinning eigenmode subspace with a mixed finite element method and applying a heterogeneity-aware cubature scheme yields a solver robust to extreme heterogeneities with performance decoupled from the resolution of the underlying mesh.
Our method provides exciting opportunities for future work. 
There exists a complex cost/quality tradeoff between dense globally supported subspaces, and sparse locally supported subspaces. Understanding this tradeoff would help resolve the artifacts  in \reffig{global_limitations} and would pave the way to robust reduced-space contact simulation -- a difficult open problem for reduced space methods \cite{lei2020medialelastics}. 
Finally, we believe our subspace MFEM solver could be extended for use in physics based inverse design in engineering and biomechanics where domain heterogeneities are commonplace.%

%% file: acknowledgements.tex
\begin{acks}
We would like to thank Aravind Ramakrishnan, Chief Trusty for proof-reading and Kevin Caldwell and Kevin Wang for providing simulation meshes. This research was made possible with the administrative help of our lab system administrator John Hancock and financial officer Xuan Dam. This research was funded by two NSERC Discovery grants, an Ontario Early Researchers Award and the Canada Research Chairs Program.
\end{acks}

%% file: extra_figures.tex
 \hspace{20 mm}

\begin{figure*}
\includegraphics[width=\textwidth,keepaspectratio]{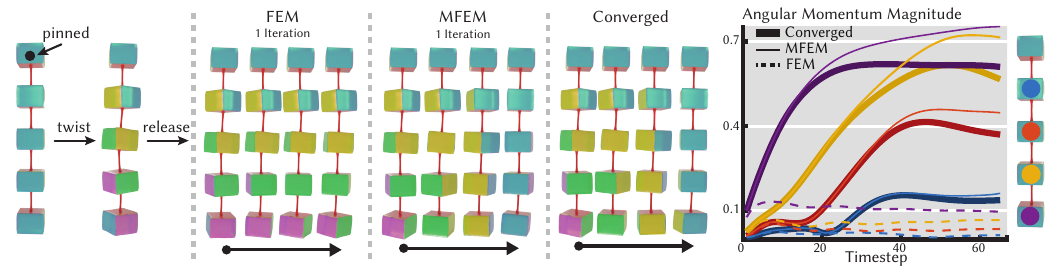}
\timestamp{\tsHeteroPendulum}
\caption{
We pin the pendulum from the top, twist the bottom end, and simulate the unwinding. We compare results from FEM and MFEM with one solver iteration per timestep against a converged subspace FEM solution. Even at low iterations our MFEM solvers show much better agreement, which is reflected on the plot on the right where total angular momentum for each pendulum block is plotted over time.
\label{fig:hetero_geometry}} 
\end{figure*}

\begin{figure}[h]
    \centering  \includegraphics{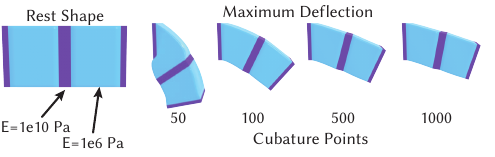}
    \caption{\Edit{Under-integration with our clustering-based integration leads to artificial softening in softer regions. Here we simulate a 48,000 tetrahedra heterogeneous cantilevered beam and visualize the maximum deflection with different numbers of cubature points. 5 skinning modes are used for this simulation.}}
    \label{fig:cubature_limitations}
\end{figure}

\begin{figure}[h]
    \centering  \includegraphics{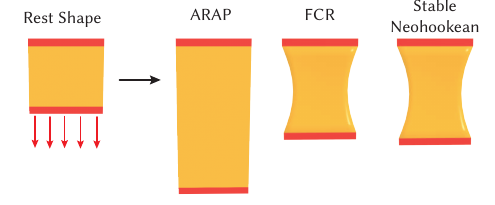}
    \caption{\Edit{Our method is compatible with any hyperelastic material model. Here we apply a load and simulate beams to equilibrium with As-Rigid-As-Possible (ARAP), fixed corotational (FCR), and stable Neohookean material models  }}
    \label{fig:material_models}
\end{figure}

\begin{figure}
\includegraphics[width=\linewidth,keepaspectratio]{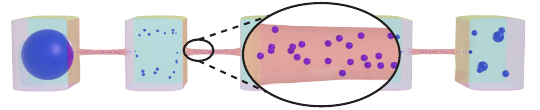}
\caption{Our cubature sampling scheme is geometry aware and constraint aware. Note that our schemes samples more densely in the thin regions and only samples a single point on the far left where the pendulum is pinned.
\label{fig:hetero_geometry_cubature_points}} 
\end{figure}